\documentclass[aip,apl,amsmath,amssymb,reprint]{revtex4-1}

\usepackage{graphicx}
\usepackage{dcolumn}
\usepackage{bm}
\usepackage{xcolor}
\usepackage{siunitx}
\usepackage[percent]{overpic}
\usepackage[utf8]{inputenc}
\usepackage[T1]{fontenc}
\usepackage{mathptmx}

\begin{document}
	

\title{Near Transform-limited Quantum Dot Linewidths in a Broadband Photonic Crystal Waveguide}

\author{Freja T. Pedersen}
\author{Ying Wang}
\author{Cecilie T. Olesen}
\affiliation{Center for Hybrid Quantum Networks (Hy-Q), Niels Bohr Institute, University of Copenhagen, Blegdamsvej 17, DK-2100 Copenhagen, Denmark}
\author{Sven Scholz}
\author{Andreas D.~Wieck}
\author{Arne Ludwig}
\affiliation{Lehrstuhl f{\"u}r Angewandte Festk{\"o}rperphysik, Ruhr-Universit{\"a}t Bochum, Universit{\"a}tsstrasse 150, D-44780 Bochum, Germany}
\author{Matthias C. L\"{o}bl}
\author{Richard J. Warburton}
\affiliation{Department of Physics, University of Basel, Klingelbergstrasse 82, CH-4056 Basel, Switzerland}
\author{Leonardo Midolo}
\author{Ravitej Uppu}
\email{ravitej.uppu@nbi.ku.dk}
\author{Peter Lodahl}
\email{lodahl@nbi.ku.dk}
\affiliation{Center for Hybrid Quantum Networks (Hy-Q), Niels Bohr Institute, University of Copenhagen, Blegdamsvej 17, DK-2100 Copenhagen, Denmark}

\begin{abstract}
Planar nanophotonic structures enable broadband, near-unity coupling of emission from quantum dots embedded within, thereby realizing ideal singe-photon sources. 
The efficiency and coherence of the single-photon source is limited by charge noise, which results in the broadening of the emission spectrum.
We report suppression of the noise by fabricating photonic crystal waveguides in a gallium arsenide membrane containing quantum dots embedded in a $p$-$i$-$n$ diode.
Local electrical contacts in the vicinity of the waveguides minimize the leakage current and allow fast electrical control ($\approx\SI{4}{\mega\hertz}$ bandwidth) of the quantum dot resonances.
Resonant linewidth measurements of $79$ quantum dots coupled to the photonic crystal waveguides exhibit near transform-limited emission over a $\SI{6}{\nano\meter}$ wide range of emission wavelengths.
Importantly, the local electrical contacts allow independent tuning of multiple quantum dots on the same chip, which together with the transform-limited emission are key components in realizing multiemitter-based quantum information processing.
\end{abstract}

\maketitle
An on-demand source of indistinguishable single photons is a key building block in a scalable quantum network\cite{kimble2008}.
Achieving on-demand operation requires high quantum efficiency of the emitter together with deterministic coupling to a single propagating mode for efficient extraction. 
Semiconductor quantum dots (QDs) coupled to nanophotonic structures have over the past decade proven to be strong candidates for such a source\cite{Melet2008,lodahl2015,Somaschi2016,fox2018,Michler2019,Anderson2020}. 
In particular photonic crystal waveguides (PCWs) enable near-unity coupling of the QD emission to a single propagating mode\cite{Arcari2014} that can be efficiently extracted across a broad spectral range.

The semiconductor environment of the QD usually introduces fluctuations in the form of charge noise, which leads to a decrease of coherence \cite{pan2016}. The charge noise stems from variations in the electronic states around the QD, which leads to fluctuations in the local electric field. 
These changes, shift the QD emission energy through the Stark effect and results in a broadening of the emission line.
Consequently, the optical linewidth increases significantly above the transform limit determined by the spontaneous emission rate \cite{Kuhlmann2013}. 
Charge noise can be suppressed by embedding the QDs in a diode heterostructure \cite{warburton2013}.
So far transform-limited QDs have been reported in bulk samples \cite{kuhlmann2015}, microcavities \cite{najer2019}, and in a multimode nanobeam waveguide \cite{lodahl2018}.
However, in many cases broadband approaches featuring efficient photon-emitter coupling is a major asset, for instance in spin-physics experiments relying on the simultaneous coupling of several optical transitions\cite{Javadi2018}.
While microcavities enable near-unity coupling efficiency, the operable spectral window is limited to the narrow linewidth of the cavity.
Multimode waveguides support broadband operation, but the couping efficiency is limited to $<90\%$.
In contrast, PCWs enable broadband near-unity coupling as well as Purcell enhancement of the QD coupled to the waveguide mode.
PCWs are composed of air holes etched in a thin membrane (see Fig. 1(a)), resulting in proximity of etched surfaces to the QDs.
The proximity of etched surfaces leads to an increase in the charge noise due to the presence of surface charge traps \cite{Liu2018Noise,Wilson2018}.
Therefore, obtaining transform-limited emission in PCWs is a challenging task and requires low-noise heterostructure semiconductor material and careful nanofabrication in addition to high-quality electrical contacts.

In this work, we establish that electrically-contacted QDs overcome charge noise and enable near transform-limited linewidth in a PCW.
Through fabrication of high-quality local electrical contacts, near-ideal diode operation was observed with a short $RC$ time-constant of $< 1\mu$s.
Linewidths of $79$ QDs with resonance frequencies distributed over $\SI{6}{\nano\meter}$ were measured using resonant transmission (RT) of a weak coherent state and a selection of them compared against the natural linewidth.
We observe that at least $65\%$ of the QDs coupled to the PCWs exhibit near transform-limited lineshapes.

\section{Device fabrication and electrical characterization}
\begin{figure}
    \centering
    \includegraphics{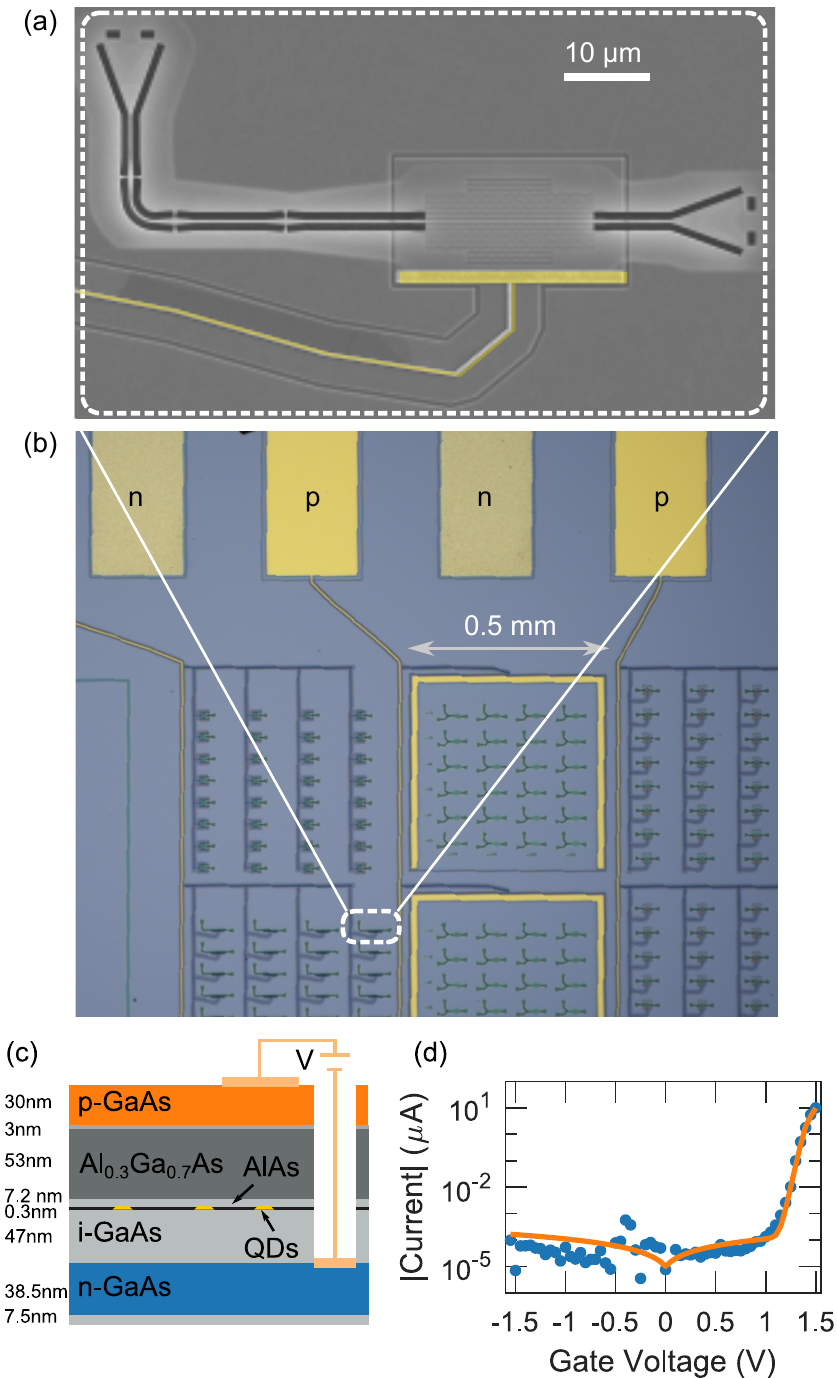}
    \caption{(a) A scanning electron microscope image of a single device with contacts and isolation trenches. (b) Optical microscope image of the fabricated device showing the arrangement of devices in individual groups connected to isolated \textit{p}-type contacts. (c) Layout of the membrane with an embedded \textit{p-i-n} diode heterostructure. (d) Current-voltage ($I$-$V$) characteristic of the \textit{p-i-n} diode measured at $T=\SI{1.6}{\kelvin}$ (circles). The solid curve is a plot of an ideal diode in series with a $R_s=\SI{7}{\kilo\ohm}$ resistor and with a finite parallel resistance of $\SI{10}{\giga\ohm}$.}
    \label{fig1}
\end{figure}

A scanning electron microscope (SEM) image of a typical device from the nanofabricated sample is shown in Fig. \ref{fig1}(a). 
The device consists of a PCW terminated with shallow-etched grating outcouplers, fabricated in a 170 nm-thin suspended gallium arsenide (GaAs) membrane as follows. 
The membrane is grown using molecular beam epitaxy on a (100) GaAs substrate.
The substrate is prepared for growth using an AlAs/GaAs superlattice followed by a $\SI{1150}{\nano\meter}$-thick Al$_{0.75}$Ga$_{0.25}$As sacrificial layer. 
A layout of the $\SI{170}{\nano\meter}$-thick GaAs membrane containing a layer of self-assembled InAs QDs grown on top of the sacrificial layer, is shown in Fig. \ref{fig1}(c). 
The QDs are located at the center of the membrane, which ensures maximal coupling of the QD emission to the transverse electric (TE) modes of the waveguide.
The membrane comprises an ultra-thin \textit{p-i-n} heterostructure diode with the layout shown in Fig. \ref{fig1}(c), which is used to apply an electric field across the QDs.
The electric field helps to reduce the charge noise and allows to tune the QD emission wavelength via the Stark effect.
The \textit{n}-type region is located $\SI{47}{\nano\meter}$ below the QDs to suppress cotunneling and at the same time stabilize the QDs charge state by Coulomb blockade\cite{Petroff1994}. 
A monolayer of AlAs capping of the QDs removes the electron wetting layer states \cite{Lobl2019}.
A $\SI{53}{\nano\meter}$-thick Al$_{0.3}$Ga$_{0.7}$As layer above the QDs is used as a blocking barrier to limit the current to a few $\si{\nano\ampere}$ at a bias voltage of $\approx \SI{1 }{\volt}$, where the QDs can be charged with a single electron.

Reactive-ion etching (RIE) in a BCl$_3$/Ar chemistry is used to open vias to the \textit{n}-layer.
The Ni/Ge/Au/Ni/Au contacts are fabricated using electron-beam physical vapor deposition followed by annealing at $\SI{430}{\celsius}$. 
To establish Ohmic \textit{p}-type contacts, Cr/Au contacts are deposited on the surface without further annealing.
The shallow-etched grating couplers are patterned by electron-beam lithography (Elionix F-125, acceleration voltage of $\SI{125}{\kilo\electronvolt}$) and then etched using RIE to a depth of approximately $\SI{50}{\nano\meter}$ \cite{zhou2018}.
The PCWs are fabricated with the process described in Ref. \citenum{midolo_soft_2015}, followed by hydrofluoric acid undercut to create suspended waveguides. 

The full processed chip has a size of \mbox{$\SI{3}{\milli\meter}\times\SI{3}{\milli\meter}$}, and is divided into five sections with physical dimensions of \mbox{$\SI{0.5}{\milli\meter}\times\SI{3}{\milli\meter}$} each. 
An optical image displaying some of these sections is shown in Fig. \ref{fig1}(b). 
Each section is connected to separate pairs of electrical contacts, also visible in the image. 
This design reduces the number of defects or thread dislocations on each diode (i.e. each section), thereby reducing the leakage current.
The \textit{n}-doped layer is used as a common ground plane for all devices, while \textit{p}-doped layers and metal wires are used to distribute the voltage uniformly to several devices in parallel. 
In order to achieve minimum cross-talk between the different sections, an isolation trench with a width of $\SI{1}{\micro\meter}$ is patterned around the \textit{p}-contacts and etched with RIE together with the shallow-etched gratings. 
Some sections are designed with local electrical contacts, such that the field can be applied to a single device. 
By bringing the contacts close to the QDs and introducing isolation trenches, the capacitance $C$ and the sheet resistance $R$ of the diodes are significantly reduced.
This reduction in the contacted area shortens the $RC$ response time of the device, and allows for fast operation of the diode.
The local electrical contact highlighted in the SEM image in Fig. \ref{fig1}(a) enables individual control of multiple devices on the same chip, which is crucial e.g. for scaling up to interfering multiple emitters\cite{Baranger2013,Mahmoodian2018}. 

The sample was cooled to $\SI{1.6}{\kelvin}$ in a closed-cycle helium cryostat with optical and electrical access for QD spectroscopy measurements.
The different sections on the sample were wired to independent twisted pair transmission lines in the cryostat.
As the \textit{n}-contacts are not isolated on the sample, we connect them to the common ground of a multichannel low-noise voltage source ($V_\textrm{rms} < \SI{1}{\micro\volt}$).
This removes potential ground loops caused by any variations in the parasitic resistance on the sample or the transmission lines.
We typically measure an RMS voltage noise of $< \SI{200}{\micro\volt}$ up to a bandwidth of $\SI{10}{\mega\hertz}$ on the sample transmission lines.
This was observed to be limited mostly by the ambient noise picked up by the twisted pair lines.
In another cryostat employing coaxial lines, RMS voltage noise of $<  \SI{80}{\micro\volt}$ has been measured on a similar sample.

The current-voltage (\textit{I-V}) curve recorded using a sourcemeter is shown in the Fig. \ref{fig1}(d).
A clear diode turn-on at gate voltage $V_g > \SI{0.7}{\volt}$ is observed with very low leakage current. 
The sample exhibits a near-ideal \textit{I-V} curve for a \textit{p-i-n} diode, with the leakage current limited by the source meter noise in the reverse bias.
In the Coulomb blockade regime for QD neutral excitons ($V_g < \SI{1.28}{\volt}$), the leakage current across the diode is $<\SI{1}{\nano\ampere}$, and has thus excellent \textit{I-V} properties.\\

\section{Resonant linewidth measurements}
\begin{figure}
    \centering
    \includegraphics{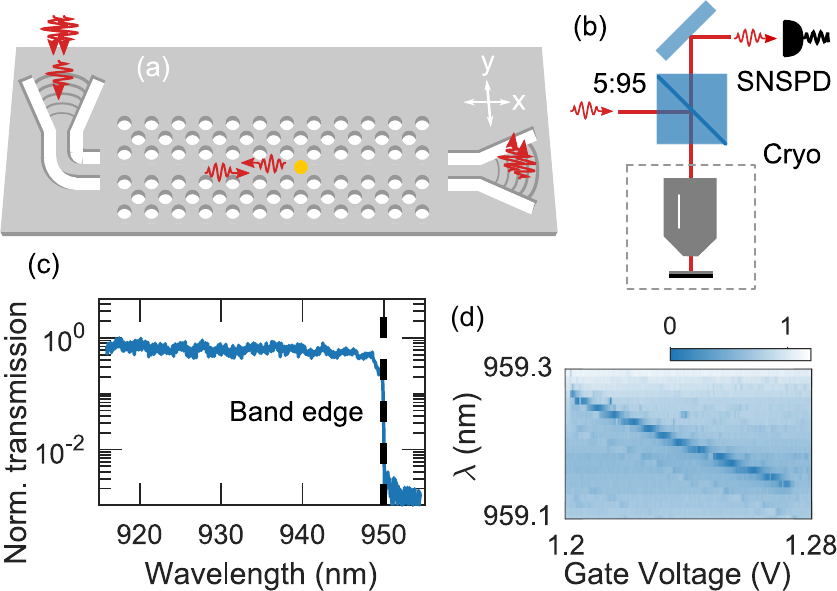}
    \caption{(a) Schematic of the photonic crystal waveguide device employed in the measurements with lattice constant $a = \SI{248}{\nano\meter}$ and hole radius $r = \SI{70}{\nano\meter}$. The photonic crystal waveguides are terminated with high-efficiency shallow-etched grating couplers, where the left grating incouples the laser and the right outcouples the transmitted signal. The QD is illustrated with a yellow dot, where a resonant photon is reflected in an RT measurement. (b) Schematic of the optical setup used for transmission measurements. (c) Frequency dependent laser transmission through the photonic crystal waveguide without QDs, normalized to transmission through a nanobeam waveguide. (d) Charge plateau of a QD neutral exciton coupled to the PCW observed in the resonant transmission of an attenuated narrow-linewidth laser.}
    \label{fig2}
\end{figure}

A schematic of the laser transmission experiment in the PCW is shown in Fig. \ref{fig2}(a). 
The  PCWs used in our measurements have a lattice constant of $a =  \SI{248}{\nano\meter}$ and hole radii $r= \SI{70}{\nano\meter}$.
The PCW is mode-matched to a section with nanobeam waveguide at both ends and terminated with high-efficiency shallow-etched grating couplers\cite{zhou2018} for in- and out-coupling of light.
Light is launched into the waveguide from the left, and light transmitted through the PCW is collected on the right grating. 
In Fig. \ref{fig2}(b) a schematic of the optical setup is shown, where a tunable narrow-band laser (bandwidth $< \SI{1}{\mega\hertz}$) is collimated and imaged to the back focal plane of a wide-field microscope objective. 
The objective focuses the laser to a spot size which is mode-matched with the in-coupling grating. 
The light transmitted through the PCW is collected from the right grating using the same microscope objective. 
The incident laser and collected transmission are separated into different spatial modes using a 5:95 (reflection:transmission) beam splitter, where the transmission arm is used for collection. 
The collected signal is detected using a superconducting nanowire single-photon detector (SNSPD).
The frequency-dependent transmission of the laser through the device at $V_g =  \SI{1.0}{\volt}$, where no QD states are populated due to the high built-in electric field, is shown in Fig. \ref{fig2}(c). 
The transmission spectrum is normalized to the transmission through a nanobeam waveguide terminated with identical grating out-couplers, to factor out the frequency-dependent diffraction efficiency of the grating out-couplers.
A steep cutoff in the transmittance of $>2$ orders of magnitude is observed at wavelengths longer than $ \SI{950.2}{\nano\meter}$, which corresponds to the band gap of the photonic crystal.
This large suppression together with a nearly constant transmission below the cutoff wavelength highlight the excellent photonic properties of the nanofabricated PCWs. 

QDs that are efficiently coupled to the PCW exhibit a single-photon non-linearity, where the single-photon component of a weak laser resonant with a QD transition is reflected as illustrated in Fig. 2(a).\cite{shen2005,javadi2015}
If the incident photon is detuned from the QD resonance, the interaction with the QD vanishes and the photon is transmitted. This resonant scattering appears as a dip in the frequency-dependent transmission across the PCW. Fig. \ref{fig2}(d) displays such a dip from an RT measurement of a QD, while tuning the applied bias voltage. 
The charge plateau shows the distinct Coulomb blockade regime for the neutral exciton.
Importantly, the charging of the QD occurs at a gate voltage close to the predicted value from bandstructure simulations in contrast to earlier reports \cite{pinotsi2011}. 
This agreement is a consequence of the low contact resistance of the sample.

\begin{figure*}
    \centering
    \includegraphics{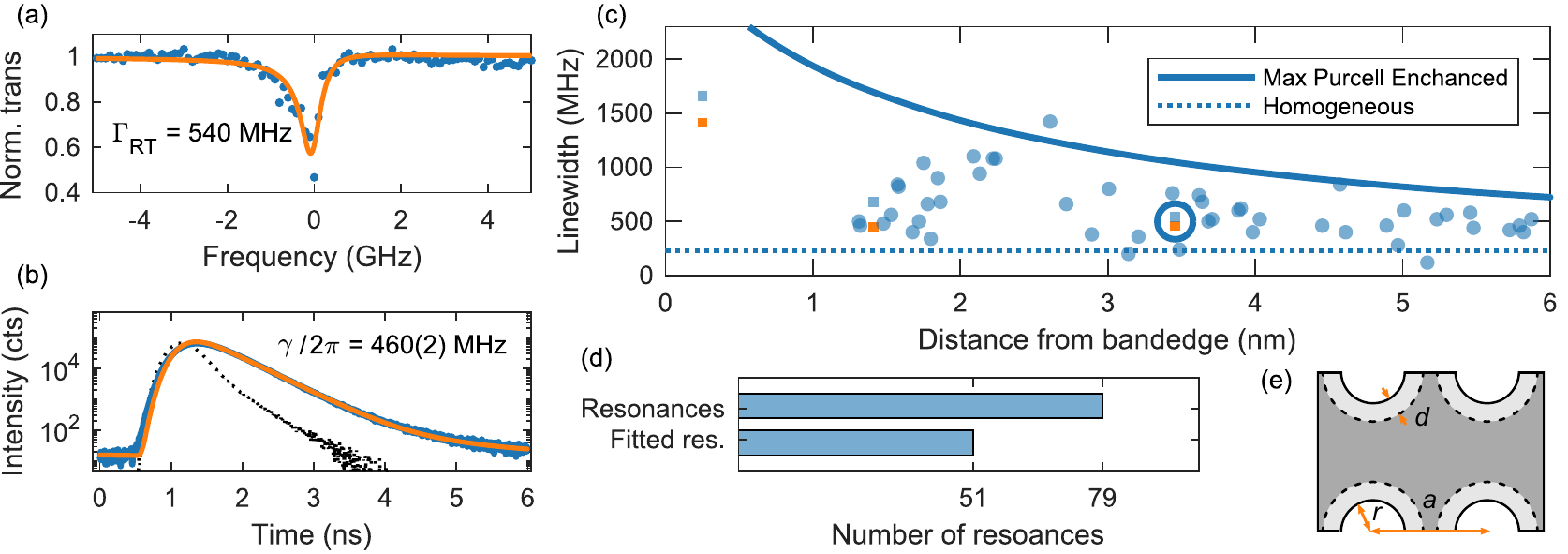}
    \caption{(a) RT line fit of a QD in the photonic crystal waveguide. The FWHM of the fitted linewidth is displayed on the plot. (b) Lifetime of the resonantly excited QD exhibiting a single exponential decay. The black dotted curve is the instrument response function (IRF) of the measurement setup. The single-exponential fit to the data includes the convolution of the model with the IRF. (c) Blue points are the FWHM of fitted RT linewidths similar to (a) for $51$ resonances of QDs in the photonic crystal waveguide. Orange squares are transform limited linewidths of three of the QD, extracted from the decay rate of lifetime measurements similar to (b). The corresponding RT linewidth are also plotted with square (blue) maker. The encircled dot are the measurements shown in (a and b). Solid blue line is the maximal achievable Purcell enhanced linewidth given the homogeneous linewidth shown with a dotted line. (d)  Overview of the total number of QD resonances found in the RT measurement and the number of QD resonances that had a pronounced RT dip whose linewidth could be extracted. (e) Photonic crystal waveguide with hole radius $r$ and lattice constant $a$. The lighter grey area at distance $d$ from the holes illustrate the region where the QDs are affected by surface charges.}
    \label{fig3}
\end{figure*}

The width of the RT dip, at powers well below the saturation power for the QD, is a reliable measurement of the QD linewidth.\cite{shen2005}
Any charge noise causing the energy levels to shift or fluctuate will degrade the photon-emitter interactions and results in a broadening of measured linewidth. 
The incident narrow-bandwidth laser power was attenuated to $P = \SI{0.4}{\pico\watt}$ in the waveguide, which was found to be $<1\%$ of the saturation power of the QDs. 
At a gate voltage of $V_g =  \SI{1.24}{\volt}$, which corresponds to populating the neutral exciton, the wavelength of the laser is scanned from $\SI{944}{\nano\meter}$ to $\SI{950}{\nano\meter}$.
The laser wavelength was locked using a wavemeter with a resolution of $\SI{50}{\mega\hertz}$ ($\SI{0.15}{\pico\meter}$).
The $\SI{6}{\nano\meter}$ wavelength range near the cutoff was chosen to capture the slow-light regime of light transport in the PCW, which leads to a Purcell enhancement in the radiative decay rate of the QD\cite{lodahl2018josab}.
Several QD resonances, identified as RT dips, were recorded in a single continuous wavelength scan, spanning the whole bandwidth of $\SI{6}{\nano\meter}$, with a step size of $\SI{100}{\mega\hertz}$ and each RT dip was independently fitted using the model described in Ref. \citenum{javadi2015}.
One such RT dip measured from a QD is shown in Fig. \ref{fig3}(a) together with the fit to the model.
The slight asymmetry in the lineshape is incorporated into the model as a Fano parameter.
The Fano lineshape of the RT dip is due to the interference between the QD resonance and a weak reflection from the mode adapters between PCW and the nanobeam waveguides in the device.
For simplicity the linewidth of the QD resonance $\Gamma_\mathrm{RT}$ is extracted as the full-width at half-maximum of the fitted curve with the Fano parameter omitted (i.e. symmetric lineshape).
This approach leads to upper-bound estimates of the linewidth.

For comparison, the transform-limited linewidth $\Gamma$ is extracted from a time-resolved resonance fluorescence (RF) measurement of the QD under pulsed resonant excitation. 
In this configuration, the QD is excited with a pulsed laser (repetition rate $= \SI{72.6}{\mega\hertz}$; pulse length $\approx  \SI{10}{\pico\second}$).
The photons emitted by the QD couple to the propagating PCW mode and are collected at the grating as before.
The spatial separation between the QD (excitation laser beam position) and the collection grating ensures a large suppression of the excitation laser in the collection.
The RF measurement involves precisely locating the QDs and suppressing the resonant laser scatter, which is challenging and time consuming.
Hence, the time-resolved measurements were carried out only on 3 of the QDs at distinct frequencies.
Furthermore, time-resolved measurements were also carried out for a few QDs outside the nanostructures (i.e. in bulk) to estimate the homogeneous linewidth of the QDs.
Figure \ref{fig3}(b) shows the time-resolved resonance fluorescence of the QD whose RT measurement is shown in Fig. \ref{fig3}(a). 
The data is modelled with a single-exponential decay convolved with the instrument response function to extract the radiative decay rate $\gamma$.
The extracted $\gamma$ is used to estimate the transform-limited linewidth $\Gamma = \gamma/2\pi =  \SI{460}{\mega\hertz}$.
This estimate of $\Gamma$ may be considered a lower bound, since any additional non-radiative processes would increase it, however non-radiative recombination was found to be negligible in photonic-crystal membranes in a previous work \cite{Wang2011}.
Comparing the natural linewidth against $\Gamma_\textrm{RT}$, we extract that $\Gamma_\mathrm{RT} \leq 1.17\,\Gamma$, which demonstrates that the noise affecting the QD is strongly suppressed. 
The residual broadening could be attributed to a slow spectral diffusion of the QD resonances (typical timescale of $>\SI{5}{\milli\second}$ in our sample) and nuclear spin noise \cite{Kuhlmann2013}.

The RT linewidths of the QDs are plotted against their spectral location with respect to the bandedge (waveguide cutoff) in Fig. \ref{fig3}(c) with blue filled markers. 
The transform-limited linewidths estimated from RF measurements are plotted as orange squares, with their corresponding RT linewidth also indicated with squares. 
The QD whose data is shown in Fig. \ref{fig3}(a-b) is marked with an open circle.
Only one QD very close to the bandedge was analyzed due to the difficulty of fitting the RT dips on the steeply rising bandedge.
The three QDs with $\Gamma$ and $\Gamma_\textrm{RT}$ measurements illustrate the near transform-limited performance of the QDs coupled to PCW over a broad wavelength range. The measured RT dips cover a large range of wavelengths, demonstrating the large bandwidth performance of the PCW.
The observed variation in the measured linewidths ($\SI{120}{\mega\hertz} - \SI{1660}{\mega\hertz}$) across the QDs is a consequence of the wavelength and spatial position dependence of the Purcell factor in a PCW.\cite{lodahl2018josab}
Using the radiative decay rate measurements of QDs outside the nanostructure, we estimate the average homogeneous linewidth to be $\Gamma_\mathrm{hom}\approx \SI{230 \pm 40}{\mega\hertz}$ (dashed line in Fig. \ref{fig3}(c)).
The Purcell factor at a specific wavelength is sensitive to the QD dipole orientation and location within the PCW.
The wavelength-dependence of the maximum Purcell enhanced linewidth is extracted from numerical calculations, and plotted as the solid curve in Fig. \ref{fig3}(c).
We observe that the maximum Purcell enhanced linewidth follows the measured $\Gamma_\textrm{RT}$ at all wavelengths as an upper-bound.
Few QDs exhibit $\Gamma_\textrm{RT}$ below the homogeneous linewidth, which indicates suppression of the radiative decay rate, and remarkably narrow linewidths are achieved.
This observation is also consistent with the PCW's ability to suppress decay rates depending on the QD dipoles location and orientation.

A total of $79$ QD resonances were found in the frequency scan on two devices, of which $51$ (i.e. $65\%$) were modeled to extract $\Gamma_\textrm{RT}$ as summarized in Fig. \ref{fig3}(d). 
The linewidths of the remaining $28$ QDs were not analyzed due to one of the two factors: (i) very shallow RT dip that could not be robustly fitted or (ii) noisy and spectrally broad RT dip that was affected by slow timescale spectral diffusion.
Since the QDs are randomly distributed in the PCW, a selection of them would be weakly coupled to the waveguide mode, either due to their spatial position or the orientation of the dipole.
Such QDs exhibit a shallow RT dip due to the weak coupling to the PCW.
We assume that the remaining probed QD resonances exhibiting noisy RT dips were influenced by surface charges causing additional spectral diffusion .
This assumption is the worst-case scenario for surface induced noise as the noise from QD growth is neglected.

We estimate an approximate upper-bound on the distance $d$ from the etched surfaces, beyond which the QDs exhibit transform-limited lineshape as follows.
Let us assume that the fraction of the QD resonances that are not modelled by a lorentzian to be an estimate for the fraction of QDs located within the distance $d$.
The total area in a unit cell of the PCW is $A_\mathrm{total}$ (total grey shaded region in Fig. \ref{fig3}(c)).
We then define $A_\mathrm{lim}$ as the area of the PCW where QDs exhibit near transform-limited linewidth (dark grey area bounded by the dotted circles with distance $d$ to the air holes).
Then, by equating the ratio of the areas to the fraction of fitted QDs $f = 51/79 = A_\mathrm{lim}/A_\mathrm{total}$, the limiting distance is estimated to be $d<\SI{43}{\nano\meter}$.
This indicates that a semiconductor heterostructure together with the high-quality electrical contacts achieve optimal operation of the QDs close to the center of the PCW, where efficient coupling to the guided mode is expected \cite{Arcari2014}.

\section{Electrical switching of quantum dots}
\begin{figure}
    \centering
    \includegraphics{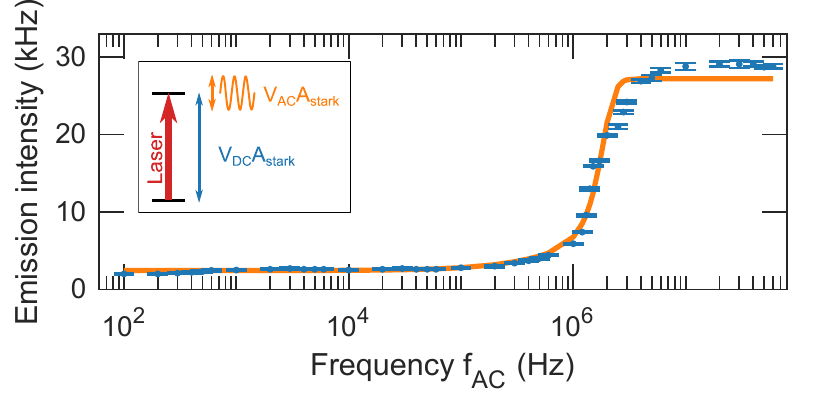}
    \caption{Measurement of the RC constant. Blue points are the cycle-averaged resonance fluorescence from a QD while sinusoidally modulating the bias voltage across the \textit{p-i-n} diode at various frequency $f_\textrm{AC}$. The orange curve is a fit to the data, which yields an RC time constant $\tau_\textrm{RC} =  \SI{0.4}{\micro\second}$.}
    \label{fig4}
\end{figure}

Alongside the noise-free operation of QDs, the near-ideal \textit{p-i-n} diode $IV$-curve with low contact resistance indicates short $RC$ time constant, which can enable fast electrical switching and control of the QD resonances \cite{prechtel2012}. 
We measure the switching time using an RF experiment, where the QD is excited by a narrow-bandwidth continuous-wave laser.
The gate voltage $V_g$ across the QD is sinusoidally-modulated around the resonant voltage of the QD.
The modulation tunes the QD in and out of resonance with the excitation laser, which in turn modulates the fluorescence intensity.
Experimentally, we employ a bias-tee to mix a DC and an AC voltage source, with the DC offset $V_\mathrm{DC}$ set to the resonant voltage of the QD and the peak-to-peak AC amplitude $V_\mathrm{AC} =  \SI{100}{\milli\volt}$, as illustrated in the inset of Fig. \ref{fig4}.

Instead of measuring the modulation in the fluorescence intensity that requires a high-frequency lock-in amplifier, we measure the time-averaged fluorescence (integration time of $\SI{1}{\second}$).
The measured fluorescence intensity with increasing frequency of the AC modulation $f_\textrm{AC}$ from $\SI{100}{\hertz}$ to $\SI{60}{\mega\hertz}$ is shown in Fig. \ref{fig4}. 
As the voltage linewidth of the QD is $ \SI{1}{\milli\volt}$, which is very small in comparison to $V_\textrm{AC}$, the QD is resonant with the laser for a very small fraction of the time per cycle.
This results in low emission intensity at small $f_\textrm{AC}$.
As $f_\textrm{AC}$ is increased close the $RC$ time constant of the diode, the amplitude of modulation experienced by the the QD decreases, which increases the emission intensity.
We observe that around $f_\mathrm{AC} =  \SI{3}{\mega\hertz}$ the Stark tuning of the QD cannot follow the $V_\textrm{AC}$ and the emission intensity saturates to the unmodulated value. 
The cycle-averaged QD fluorescence intensity $I_\textrm{QD} (f_\textrm{AC})$ is modeled as
\begin{equation}
I_\textrm{QD} (f_\textrm{AC}) = I_0 \int_{-A(f_\textrm{AC})}^{A(f_\textrm{AC})} S(V-V_\mathrm{DC}) dV,
\end{equation}
where, $A(f_\textrm{AC}) \equiv (V_\textrm{AC}/2) \exp [-2\pi f_\textrm{AC} \tau_\textrm{RC}]$, $S(V-V_\mathrm{DC})$ is the measured voltage response of the QD and $I_0$ is the measured resonance fluorescence intensity without modulation.
The model yields an $RC$ time constant $\tau_{RC} = \SI{0.4}{\micro\second}$, i.e. a cutoff frequency of $1/(2\pi\tau_{RC}) = \SI{3.98}{\mega\hertz}$ for the experimental data.
The measured $\tau_{RC}$ together with the $\SI{7}{\kilo\ohm}$ series resistance estimated from the $I$-$V$ curve results in a capacitance $C = \SI{57}{\pico\farad}$, which is close to the expected value for the contacted area \cite{midolo2017}.
The switching speed can be further increased by reducing the contacted area. \cite{pagliano2014}\\

\section{Conclusion}
In summary, we have demonstrated that charge noise in QDs coupled to PCWs can be largely overcome by employing electrical-contacted QD heterostructures.
Statistics on several QDs reveal that $65\%$ of the QDs exhibit near transform-limited linewidths, which highlight the possibility to realize scalable single-photon sources for quantum information processing \cite{PCWsource}.
While some residual surface-induced noise persists, the region of influence of this noise is limited to $<\SI{43}{\nano\meter}$ from etched surfaces and could be potentially overcome through surface passivation techniques\cite{favero2017,arakawa2019} or through deterministic positioning of QDs \cite{Hofling2017,pregnolato2019,Pascale2020}.
High-speed operation of the \textit{p-i-n} diode with a cutoff frequency of $\approx 4$ MHz is demonstrated by measuring the fluorescence from a voltage-modulated QD.
The short $RC$ time constant in combination with near transform-limited linewidths of several QDs paves the way for multi-emitter based quantum information processing protocols, which will greatly benefit from the independent and deterministic control of individual transform-limited emitters\cite{Sandoghdar2002,Lukin2016,Waks2018}.\\ 

\begin{acknowledgements}
We gratefully acknowledge financial support from Danmarks Grundforskningsfond (DNRF) (Center for Hybrid Quantum Networks (Hy-Q, DNRF139)), H2020 European Research Council (ERC) (SCALE), Styrelsen for Forskning og Innovation (FI) (5072-00016B QUANTECH), Bundesministerium f\"{u}r Bildung und Forschung (BMBF) (16KIS0867, Q.Link.X), Deutsche Forschungsgemeinschaft (DFG) (TRR 160), SNF (Project No. 200020\_156637) and NCCR QSIT. 
\end{acknowledgements}

\providecommand{\noopsort}[1]{}\providecommand{\singleletter}[1]{#1}%

\end{document}